\documentclass[a4paper,10pt]{article}
\usepackage[affil-it]{authblk}
\usepackage{graphicx}
\usepackage{color} 
\usepackage{amsfonts}
\usepackage{amsmath}
\RequirePackage{fix-cm}
\usepackage{multirow}
\usepackage{multicol}
\usepackage{array}
\usepackage[paperheight=25cm,left=2.2cm,right=2.2cm,top=2.0cm, bottom=2.5cm]{geometry}

\newcommand{\ct}{\cite}
\newcommand{\bi}{\bibitem}
\newcommand{\be}{\begin{equation}}
\newcommand{\ee}{\end{equation}}
\newcommand{\ba}{\begin{eqnarray}}
\newcommand{\ea}{\end{eqnarray}}

\begin{document}
\title{Quantum Annealing in Sherrington-Kirkpatrick Spin
Glass in Presence of Time-Dependent Longitudinal
Field}    

%\author{Atanu Rajak}\email{atanu@phy.iith.ac.in}\affiliation{Department of Physics, Indian Institute of Technology, Hyderabad 502284, India}
%\author{Bikas K Chakrabarti}\email{bikask.chakrabarti@saha.ac.in}\affiliation{ Saha Institute of Nuclear Physics, Kolkata 700064, India}

\author{Atanu Rajak%
  \thanks{Electronic address: \texttt{atanu@phy.iith.ac.in}}}
    \affil{Department of Physics, Indian Institute of Technology, Hyderabad 502284, India}
  \author{Bikas K Chakrabarti%
  \thanks{Electronic address: \texttt{bikask.chakrabarti@saha.ac.in}}}
\affil{Saha Institute of Nuclear Physics, Kolkata 700064, India}

\maketitle
\begin{abstract}
\noindent Motivated by the recent development of quantum technology using quantum annealing technique and the recent works on the static properties of the Sherrington-Kirkpatrick (SK) spin glass model, we study quantum annealing of the spin glass model by tuning both transverse and longitudinal fields. We numerically solve the time-dependent Schr\"odinger equation of the total Hamiltonian when both the fields are made time-dependent and eventually vanish at the same time. We have computed the time-evolution of the probability of finding the system in one of two degenerate ground states of the classical spin glass. At the end of annealing, using the configuration averaged probability,  we have shown a clear advantage while the longitudinal field is annealed rather than keeping it constant throughout the process of quantum annealing. We further investigate the order parameter distribution of a quantum SK spin glass in presence of a small longitudinal field and find, from our exact diaginalization results
for small system sizes, evidence for quantum
tunneling induced disappearance of the classical
Almeida-Thouless phase boundary separating the
replica symmetry broken (nonergodic) and replica
symmetric (ergodic) spin glass phase (reported
already in $2022$). We believe that this longitudinal
field induced ergodicity in quantum SK model to
be responsible for the observed enhancement of
quantum annealing (reported earlier for smaller
systems by us in $2014$).
%\textcolor{blue}{and find the replica symmetric (ergodic) spin glass phase even for systems with small size compared to the case of zero longitudinal field.}
%and indicate the speeding up of the ergodicity as compared to the zero longitudinal field case. 
%Our speculation is that this emergent ergodicity is responsible for the advantage in quantum annealing with annealed longitudinal field.
%\textcolor{blue}{We believe that this emergent ergodicity is responsible for the advantage in quantum annealing with annealed longitudinal field.}
\end{abstract}
\textbf{$~~~~~~$Keywords:} Quantum annealing; Spin glass; Quantum tunneling
\vskip 0.2cm
%\textbf{$~~$PACS Nos.:} 03.67.Lx; 75.10.Nr; 75.45.+j

\section{Introduction}
\label{intro}
%Quantum annealing (QA) is a generic tool to solve combinatorial optimization problems using quantum tunneling~\ct{kadowaki98,das08,rajak22}. 
Quantum annealing (QA) is a generic tool to find near-optimal solutions of optimization 
problems using quantum tunneling~\ct{kadowaki98,das08,rajak22}.
In recent years, D-wave systems has developed and commercialized programmable QA machines known as quantum annealers using the idea of QA~\ct{johnson11}, leading to an upsurge to study QA as a novel approach of adiabatic quantum computation. 
%QA usually attempts to find the ground state of a generic Ising model having many-body random interactions. 
QA usually attempts to find the solutions of general optimization problems that can be 
mapped to a problem in searching the ground state of a generic Ising model having 
many-body random interactions.
Therefore, the applications of QA are not limited to only quantum computing but many daily life problems like finance~\ct{venturelli19}, marketing~\ct{nishimura19}, traffic flow~\ct{neukart17}, the traveling salesman problem~\ct{martonak04}, and so on, thus attracting much attention from science to the industrial community. It can be shown that these problems are effectively reduced to finding out the ground state of a random Ising model, justifying the reason for such wide applications of QA. 

%The role of quantum tunneling which is
%the key ingredient in QA was first
%discussed 
The quantum tunneling as the key ingredient of QA was first discussed by Ray et al.~\ct{ray89}, in 1989 in
the context of the restoration of the
replica symmetry and consequent ergodicity
in quantum Sherrington-Kirkpatrick (SK)
spin glass model due to quantum
fluctuations. 
Indeed since then, many later studies identify QA as a
"technique inspired by classical simulated annealing that
aims to take advantage of quantum tunnelling"~\ct{boxio16} , and
after the development of the D-wave machine ~\ct{johnson11}, it is
generally believed that quantum tunneling can help
to escape from the
local minima and approach the global
minimum (ground state) of the (free)
energy landscape if the minima are
separated by (classically inescapable)
very high but narrow barriers.
The time-dependent Hamiltonian to perform QA is given by
\be
H(t)=H_0+A(t)H_D,
\label{th}
\ee
where $H_0$ is the time-independent classical many-body Hamiltonian whose ground state is to be found, and $H_D$ is the driver Hamiltonian that does not commute with $H_0$. Thus $H'(t)=A(t)H_D$ introduces quantum tunneling between different configurations of the classical Hamiltonian $H_0$ by mixing its various stationary states. 
%The time-dependent parameter $A(t)$ denotes the strength of tunneling that is reduced and eventually vanishes at large times.
For adiabatic computation~\ct{albash18}, the time-dependent
parameter $A(t)$ denotes the strength of tunneling
that is reduced and eventually vanishes at large
times.

The dynamics of the system is governed by the time-dependent Schr\"odinger equation for the total Hamiltonian
\be
i\hbar\frac{\partial|\psi(t)\rangle}{\partial t}=H(t)|\psi(t)\rangle,
\label{wf1}
\ee
where $|\psi(t)\rangle$ is the instantaneous wave function of the total Hamiltonian $H(t)$. For our calculations, we consider $\hbar=1$ without any loss of generality. As mentioned before, the driving part $H'(t)$ introduces quantum fluctuations in the system which leads to visiting the system different eigenstates (configurations) of the classical Hamiltonian $H_0$ with finite probabilities. In starting, the strength of the driving part ($H'$) of the total Hamiltonian ($H$) is kept much higher than the classical part ($H_0$) and the dynamics is indeed determined by the quantum fluctuation term. As the dynamics proceeds the tunneling strength $A(t)$ is reduced adiabatically from a high value to zero as $t\rightarrow\infty$. According to the adiabatic theorem~\ct{moitra08}, if we consider the initial state as the ground state of the total Hamiltonian at $t=0$, the instantaneous state ($|\psi(t)\rangle$) will be the ground state of the Hamiltonian at that parameter values and at the end of annealing the system will reach the ground state of the classical Hamiltonian with a finite probability. We can anticipate this result even for disordered systems like the spin glass systems whose different configurations are separated by high but narrow energy barriers.

Simulated annealing (SA) is the classical counterpart of QA, which is a well known method to solve optimization problems~\ct{kirkpatrick83}. In this method, initially the system is prepared in an arbitrary configuration and then evolves according to the Monte-Carlo method that utilizes the Boltzman probability to hop to another configuration, and anneal the system down to zero temperature. If the temperature is tuned with a sufficiently small rate, it is expected that the system will attain the global minimum or ground state that may be degenerate. SA uses thermal fluctuations to go from one local minimum to another separated by an energy barrier and the corresponding rate is given by $e^{-h/k_BT}$, where $h$, $k_B$ and $T$ are the barrier height, the Boltzman constant and the temperature, respectively. This suggests that if the barrier height is proportional to the system size $N$, an exponentially long time in $N$ is needed to reach the global minimum by the SA. In contrast, the quantum tunneling probability is approximately given by $e^{-w\sqrt{h}/g}$, where $w$ is the width of the barrier and $g$ is the strength of quantum fluctuation~\ct{mukherjee15}. Assuming $w$ as a constant and the height $h\sim O(N)$, the time necessary to achieve the ground state employing QA is proportional to exponential of $\sqrt{N}$. Due to this $\sqrt{N}$ advantage, for systems with barrier heights $h\sim O(N)$, the QA is a potential method in searching the global minimum over the SA. 

Spin glasses are the systems that have quenched disorder in terms of ferromagnetic and antiferromagnetic interactions between spins at different lattice sites, thus inducing frustration in such systems since all the interactions can not be satisfied simultaneously. This leads to a large number of local minima with a global minimum (may be degenerated) in the energy landscape which are separated by the energy barriers that are sometimes proportional to the system size ($N$). It can be justified by the fact that one needs to flip a finite fraction of $N$ spins to transit from one local minimum to another. As a consequence, finding the global minimum or ground state for such systems is not an easy task considered to be a hard problem, and it becomes NP-hard for SK spin glass model~\ct{sherrington75}. Therefore, the SA is not an efficient way to find the ground state for such systems, whereas it is shown that the QA can be an alternative route in this context. 

In present, SK spin glass systems have attracted considerable attention 
both in theory and experiments due to the recent development of D-wave machine that performs QA of various Ising spin systems. In addition, although the static properties of the quantum SK model are studied studied extensively both numerically and analytically for a few decades, there are still many ambiguities due to its enormous complex nature~\ct{mukherjee15_cla,yan15,mukherjee18,leschke21,schindler22}. Recently it is reported that a non-zero longitudinal field destroys the spin glass transition for quantum SK spin glass model~\ct{schindler22}. In this context, we are interested here to perform a QA study of the system in Eq.~(\ref{ham1}) to investigate the effect of longitudinal field. We consider the time dependences of $\Gamma$ and $h$ such that they have large initial values, and eventually going to zero at some annealing (quenching) time $\tau$. In a seminal paper, the authors have studied QA by varying the transverse field, while introducing a small value of longitudinal field so that it breaks the trivial Ising degeneracy and tends the system in one of the two degenerate ground states in the end of annealing~\ct{kadowaki98}. In our previous work, we provide an indication of benefits of the QA by varying both transverse and longitudinal fields over the QA by varying only the transverse field with a small but constant longitudinal field for a small system of $N=8$ spins~\ct{rajak14}. We now extend our study for bigger system sizes with increased number of disorder realizations. In addition, we compare our results with the cases of constant longitudinal field and indicate the advantages of our annealing schedule as we increase the system sizes. We also investigate the order parameter distribution of quantum SK spin glass both in vanishing and non-vanishing limits of the longitudinal field. Due to the rugged nature of free energy landscapes of spin glasses, one finds a broad order parameter distribution in spin glass phase which is indeed the manifestation of replica symmetry breaking or non-ergodic behavior. One can then expect that the broad distribution converges to a delta function like when the system is made ergodic by perturbing it someway. We show that if we incorporate a small value of the longitudinal field in the quantum SK model, the system becomes near ergodic even for small systems and we anticipate that it will be fully ergodic in the thermodynamic limit. This may play an important role for the advantage of QA by varying both transverse and longitudinal fields over only varying the transverse field with a constant but small longitudinal field.

The paper is organized as follows: In Sec.~\ref{model}, we introduce the Hamiltonian of SK spin glass model in presence of both time-dependent transverse and longitudinal fields and discuss the annealing schedule. In Sec.~\ref{results}, we present our numerical results on QA and order parameter distribution, and discuss the advantage of the annealing technique adopted here. Finally we conclude in Sec.~\ref{conclu}.

\section{Model and annealing schedule}
\label{model}
The Hamiltonian of the SK spin glass in presence of transverse and longitudinal fields is given by
\be
H(t)=-\sum_{i<j}J_{ij}\sigma_i^z\sigma_j^z-\Gamma(t)\sum_i\sigma_i^x-h(t)\sum_i\sigma_i^z,
\label{ham1}
\ee
where $\sigma_i^z$ and $\sigma_i^x$ are respectively $z$ and $x$ components of the Pauli spin-$1/2$ operator, and
$J_{ij}$ are independent random variables denoting the interactions between two spins at different sites. 
The random variables are drawn from a Gaussian distribution of zero mean and variance $J^2/N$, 
$\rho(J_{ij})=(\frac{N}{2\pi J^2})\exp(\frac{-NJ_{ij}^2}{2J^2})$. The first part of the total Hamiltonian (\ref{ham1}),
denoting as $H_0=\sum_{i<j}J_{ij}\sigma_i^z\sigma_j^z$, represents the Hamiltonian for the classical S-K spin glass.
The time-dependent transverse field $\Gamma(t)$ acts as the tunneling amplitude between different configurations of 
the classical spin glass. The parameter $h(t)$ is the time-dependent longitudinal field that breaks Ising spin 
symmetry of the system.

Our aim here is to find the ground state of the
classical SK spin glass (represented by the
Hamiltonian $H_0$) using the method of quantum annealing in
presence of an additional time-dependent longitudinal
field that eventually vanishes along with the
transverse field.
We consider linear annealing protocols for both transverse and longitudinal fields, defined by $\Gamma(t)=\Gamma_0(1-t/\tau)$ and $h=h_0(1-t/\tau)$ respectively, where $\Gamma_0$ and $h_0$ are the initial values of the fields at $t=0$ and $\tau$ is the annealing time. At $t=\tau$, both the fields vanish completely and the remaining part describes the Hamiltonian of classical spin glass in Eq.~(\ref{ham1}). We solve numerically the time-dependent Schr\"odinger equation (\ref{wf1}) using the Runge-Kutta method incorporating the annealing schedules in Eq.~(\ref{ham1}). The initial state for the solution is found by exactly diagonalizing the Hamiltonian (\ref{ham1}) for initial parameter values. We compute the instantaneous probability, $P(t)=|\langle\psi_0|\psi(t)\rangle|^2$ that describes the probability of finding the system in one of the ground states 
$|\psi_0\rangle$ of the classical
spin glass Hamiltonian $H_0$ at any time $t$. We performed the same analysis in our previous work, but this time with a different annealing schedule and larger system sizes.

\begin{figure}
\centering
\includegraphics[height=2.0in]{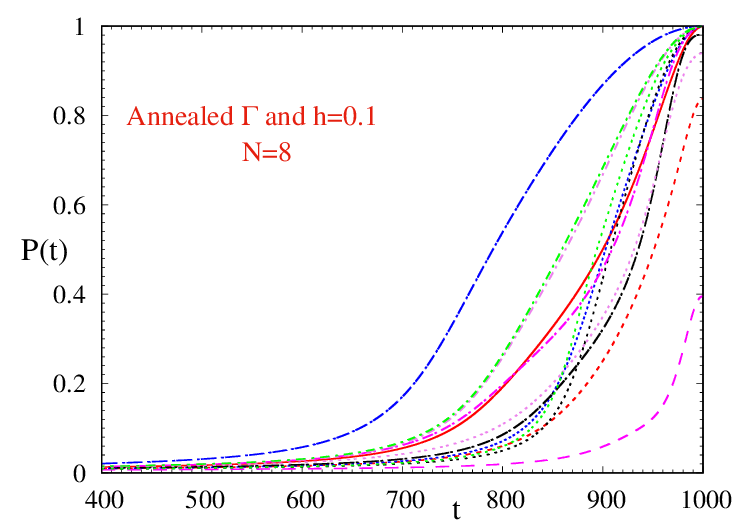}
\includegraphics[height=2.0in]{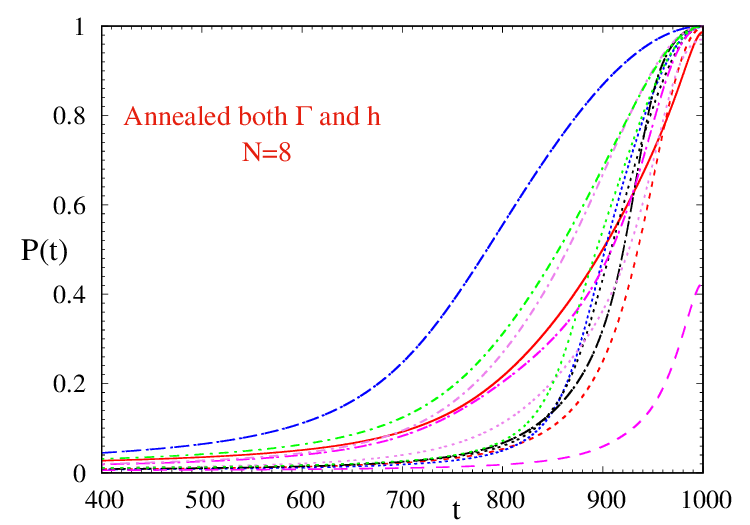}
\caption{Left panel: Time-evolution of the probability of finding the system in the classical SK spin glass 
ground state following the annealing schedule $\Gamma(t)=4.0(1-t/\tau)$, with a small constant longitudinal
field $h=0.1$ for $12$ different realizations of exchange interactions. Right panel: The same plot with the 
longitudinal field down to zero according to the relation, $h(t)=1.0(1-t/\tau)$ in addition to $\Gamma(t)=4.0(1-t/\tau)$ 
for identical sets of exchange interactions as the left panel. Here $N=8$ and $\tau=1000$.}
\label{fig_N8}
\end{figure}

\begin{figure*}
\centering
\includegraphics[height=2.0in]{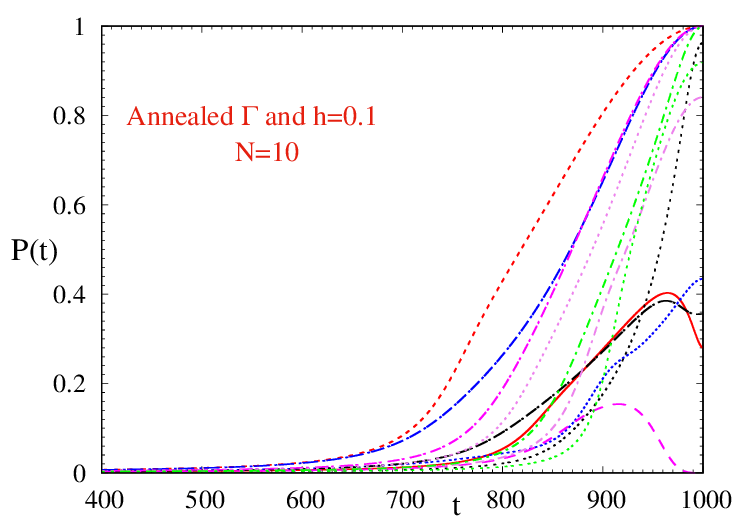}
\includegraphics[height=2.0in]{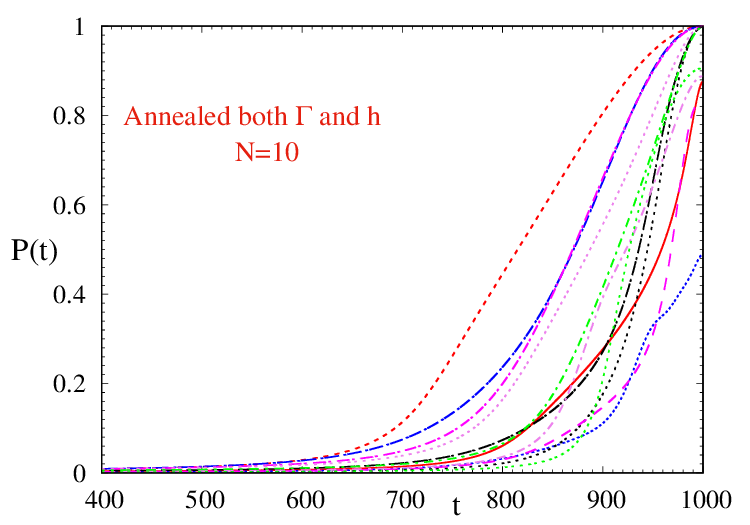}
\caption{Left panel: Variation of the probability of finding the system in the classical SK spin glass 
ground state following the annealing schedule $\Gamma(t)=4.0(1-t/\tau)$, with a small constant longitudinal
field $h=0.1$ for $12$ different realizations of exchange interactions. Right panel: The same plot with the 
longitudinal field annealed to zero according to the relation, $h(t)=1.0(1-t/\tau)$ in addition to $\Gamma(t)=4.0(1-t/\tau)$ 
for identical sets of exchange interactions as the left panel. Here $N=10$ and $\tau=1000$.}
\label{fig_N10}
\end{figure*}

\begin{figure*}
\centering
\includegraphics[height=2.0in]{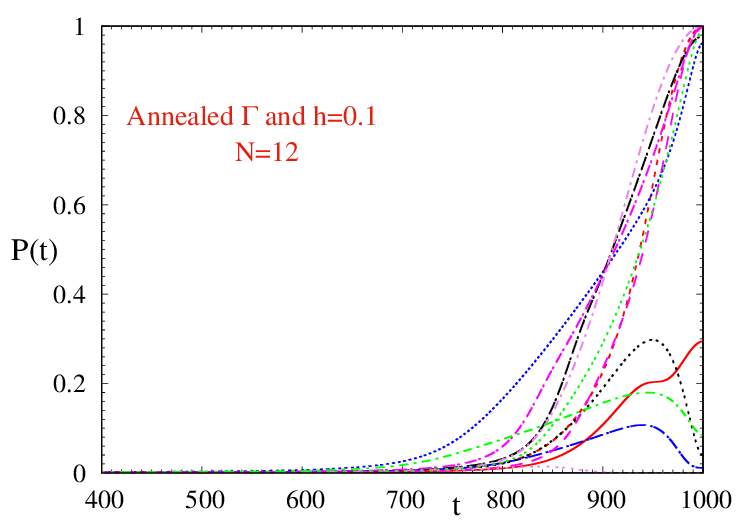}
\includegraphics[height=2.0in]{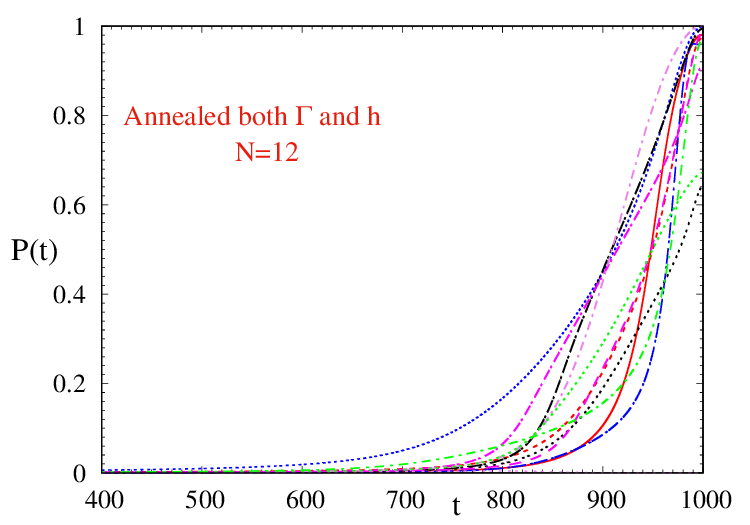}
\caption{Left Panel: The instantaneous probability $P(t)$ to find the system in the classical SK spin glass ground state following the annealing schedule $\Gamma(t)=4.0(1-t/\tau)$, with a small constant longitudinal
field $h=0.1$ for $12$ different realizations of exchange interactions. Right Panel: The same plot with the 
longitudinal field annealed to zero according to the relation, $h(t)=1.0(1-t/\tau)$ in addition to $\Gamma(t)=4.0(1-t/\tau)$ 
for identical sets of exchange interactions as the left panel. Here $N=12$ and $\tau=1000$.}
\label{fig_N12}
\end{figure*}

\section{Numerical results}
\label{results}
We now present our numerical results with the annealing schedules mentioned in Sec.~\ref{model}. The initial values of the transverse and longitudinal fields are considered as $\Gamma_0=4.0$ and $h_0=1.0$ at $t=0$. As mentioned before, we consider two annealing schemes, the first one by varying only $\Gamma$ with a small but constant $h=0.1$ and the second one by varying both $\Gamma$ and $h$. We perform our numerical analysis for three different system sizes $N=8, 10$ and $12$ with the annealing time $\tau=1000$. The exchange interactions ($J_{ij}$) are taken from a Gaussian distribution with zero mean and standard deviation $J/\sqrt{N}$ by fixing $J=1$. For each system size, we have shown the variation of $P(t)$ with time for two annealing protocols side by side (see Figs.~\ref{fig_N8},\ref{fig_N10},\ref{fig_N12}). The different curves in a single plot indicate the evolution of $P(t)$ for $12$ different realizations of random variable $J_{ij}$. For all the plots, we take overlap of the
instantaneous state $|\psi(t)\rangle$ with one of two
degenerate classical ground states of the
Hamiltonian $H_0$ that provide highest probability in the end of annealing process.
From all these plots, there is a clear indication of better performance of the annealed $h(t)$ case over the constant $h$ one in the context of overlap probability. Below we provide a quantitative measure of the probability $P(t)$ for a comparative study.

To show the advantage of QA for the $h$-annealed case over the constant but $h=0.1$ scenario, 
we consider three measures found from the evolution of $P(t)$ with time. The results are summarized in Table~\ref{table_result}. For each case of the annealing, we find the number of cases of disorder realizations out of $12$ for which $P(\tau)\approx1.0$ and also $P(\tau)\ge0.95$. In addition, we have also calculated the average of final overlap probability at $t=\tau$ over $12$ disorder realizations for all three system sizes. To evaluate the average probability ${\bar P}(t=\tau)$, we add the average probabilities found from the overlap of the instantaneous wave function with both degenerate ground states of the classical spin glass, arising due to the Ising symmetry. From all these measures, there is a clear indication of substantial advantage of the $h$-annealed case over the constant $h$ one. We note that at least for the system sizes $N=10$ and $12$, there are a few cases for which the instantaneous probability $P(t)$ initially increases and then starts decreasing after a time for $h=0.1$, whereas $P(t)$ are monotonically increasing function of time $t$ for the $h$-annealed cases. This indicates that the dynamics gets affected if we keep a constant $h$ throughout the process even though its value is small.

\begin{table}
\centering
\begin{tabular}{|m{1.8cm}|m{1.8cm}|m{1.8cm}|m{1.8cm}|m{1.8cm}|m{1.8cm}|m{1.8cm}|}
%\begin{tabular}{|l|l|l|l|l|l|l|}
\hline
\multirow{2}{*}{System Size} &
\multicolumn{2}{c|}{\parbox{4cm}{No. of disorder realizations with $P(\tau)\approx1.0$}} &
\multicolumn{2}{c|}{\parbox{4cm}{No. of disorder realizations with $P(\tau)\ge0.95$}} &
\multicolumn{2}{c|}{Average Probability ${\bar P}(\tau)$} \\
 %System Size & $P(t)\approx1.0$ & $P(t)\ge9.5$ & Average $P(t=\tau)$ \\
 \cline{2-7}
 & $h=0.1$ & $h$-annealed & $h=0.1$ & $h$-annealed & $h=0.1$ & $h$-annealed \\
\hline
$N=8$ & $8$ & $9$ & $9$ & $11$ & $0.9387$ & $0.9824$\\
\hline
$N=10$ & $5$ & $7$ & $6$ & $7$ & $0.7416$ & $0.9465$\\
\hline
$N=12$ & $4$ & $3$ & $7$ & $8$ & $0.6145$ & $0.8412$\\
%\multirow{2}{*}{\parbox{1.8cm}{\begin{flushleft}Marginal localization\end{flushleft}}} & $\phi_0=0,~~~~$  $\sigma=1.0$ & $K\lesssim\frac3{\log(t_f)}$ & $K\gtrsim\frac6{\log(t_f)}$ & 3, 4\\
%\cline{2-5}
%& $\phi_0=\pi,~~~~$ $\sigma=0.1$ & $K\lesssim\frac{1.5}{\log(t_f)}$ & $K\gtrsim\frac6{\log(t_f)}$ & 5, 6, 8 \\
\hline
\end{tabular}

\caption{Comparison of the probabilities at $t=\tau=1000$ between $h=0.1$ and $h$-annealed cases using the results in Figs.~\ref{fig_N8},\ref{fig_N10},\ref{fig_N12}. The $2$-nd and $3$-rd columns provide the number of disorder realizations having $P(\tau)\approx1.0$ and $P(\tau)\ge0.95$, respectively, for both $h=0.1$ and $h$-annealed cases with three different system sizes. The last column provides the average $P(\tau)$ over $12$ disorder realizations for both $h=0.1$ and $h$-annealed cases with three different system sizes.}
\label{table_result}
\end{table}

%To get an intuitive picture of our QA results in presence of longitudinal field, we evaluate the distribution of local order parameters of quantum SK spin glass for some fixed values of $\Gamma$ and $h$ using exact diagonalization of the Hamiltonian in Eq.~(\ref{ham1}). 
To get an intuitive picture of our QA results in
presence of longitudinal field, we first note that
the Variational Ansatz study of Schindler et al.~\ct{schindler22} for the ground state of the quantum SK model
with time-independent $h > 0$ in Hamiltonian (\ref{ham1})
has convincingly showed that the Almeida–Thouless
transition line (see e.g., \ct{binder86}) between Parisi's
Replica Symmetry Broken (non-ergodic) SK spin
glass phase and the Replica Symmetric (or ergodic)
SK phase of the classical SK model ($\Gamma = 0$
in Hamiltonian (\ref{ham1})) disappears due to quantum
fluctuations. This explains why we observed (in
~\ct{rajak14} and more clearly in this study) faster
quantum annealing in the SK glass for annealing
both the transverse and longitudinal fields by
utilizing the ergodicity (or replica symmetry)
of the system in presence of non-zero
longitudinal field until the end of annealing
process. To make it more explicit, we evaluate
here the  distribution of local order parameters
of quantum SK spin glass for some fixed values of
$\Gamma$ and $h$ using exact diagonalization of
the Hamiltonian in Eq.~(\ref{ham1}).
The Hamiltonian can be written in the matrix form using the spin up/down basis states for $N$ number of spins which are also eigenbasis states of the Hamiltonian $H_0$. Following the exact diagonalization, the $n$-th eigenstate of the Hamiltonian in Eq.~(\ref{ham1}) can be written as $|\psi_n\rangle~= \sum_{\alpha=0}^{2^{N}-1} c_{\alpha}^n |\varphi_\alpha\rangle$, 
where $c_{\alpha}^n=\langle\varphi_{\alpha}|\psi_n\rangle$ with $\sum_{\alpha}|c^n_{\alpha}|^2=1$ for any arbitrary $n$, and $|\varphi_\alpha\rangle$ denote the eigenstates of the Hamiltonian $H_0$. Since we are interested to investigate the behavior of the system at zero temperature, we will consider ground state ($|\psi_0\rangle$) averaging of different observables of interest. The order parameter for this zero temperature system can be defined as
$Q = (1/N) \sum_i \overline{\langle\psi_0|\sigma_i^z|\psi_0\rangle^2}=(1/N)\sum_i\overline{Q_i}$. The $Q_i$ can be treated as the local order parameter of the system corresponding to the $i$-th spin. The overhead bar indicates the averaging over different realizations of disorder. Then the distribution of the local order parameters can be calculated as 
\begin{equation}
P(|Q|)=\overline{\frac{1}{N}\sum_{i=1}^N\delta(|Q|-Q_i)}.
\end{equation}

We investigate here the behavior of the distribution $P(|Q|)$ for a particular value of $\Gamma$ by varying $h$ from zero to some non-zero values. To calculate the order parameter numerically, one must take care of the symmetries of the system. Our system has $Z_2$ symmetry ($\sigma^z\rightarrow-\sigma^z$) in absence of the longitudinal field $h$, whereas it breaks in the presence of the same. Therefore while calculating the local order parameter for $h=0$, we consider summing over half of the basis states, otherwise it will always give $\langle \sigma^z_i\rangle=0$ as the other half is the mirror image of the first half. To get proper values we then rescale the order parameter. However, for a non-zero $h$, we can extend the summation over all the basis states, since $Z_2$ symmetry is not preserved in that case.

\begin{figure}
\centering
\includegraphics[height=2.0in]{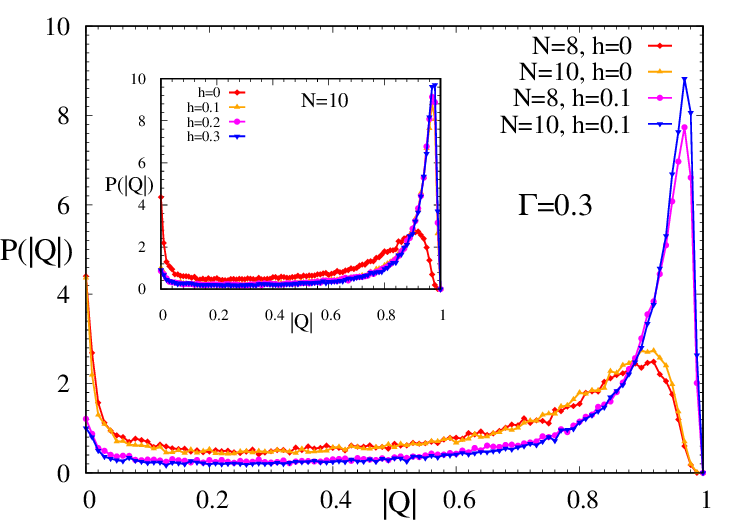}
\caption{Exact diagonalization results of distribution of local order parameters ($|Q|$) of SK spin glass in presence of a transverse field with both zero and non-zero longitudinal fields at zero temperature. For all the plots, the transverse field is fixed at $\Gamma=0.3$. In the main figure, the distribution is plotted for two different system sizes $N=8$ and $10$, with $h=0$ and $0.1$. In the inset, $P(|Q|)$ is plotted for $N=10$, and different values of $h$. For all the cases, the area under the $P(|Q|)$ curves is normalized to $1$.}
\label{fig_op_dist}
\end{figure}

We have shown the variation of $P(|Q|)$ as a function of $|Q|$ in Fig.~\ref{fig_op_dist} for $\Gamma=0.3$ for different values of $h$ with $N=8$ and $10$. We observe a broad order parameter distribution with a peak at some non-zero value of $|Q|$ and extending its tail to the zero value of the order parameter for $h=0$. We show the distributions for two system sizes in the case of $h=0$ and expect that the peak at zero order parameter vanishes for thermodynamically large systems as shown in one of our previous works~\cite{mukherjee18}. Remarkably, the scenario changes drastically if we introduce a small value of $h$ even for small systems studied here. The peak at $|Q|=0$ gets reduced substantially, whereas the peak at non-zero value of $|Q|$ becomes very sharp as compared to the $h=0$ cases. In the inset of Fig.~\ref{fig_op_dist}, we consider various values of $h$ for a system of size $N=10$ to show the stability of our results with respect to the application of $h$. The peak at non-zero value of $|Q|$ grows with increasing $h$ for a single system size. 
%This provides a clear indication of speed up of ergodicity of quantum SK model in presence of a small but finite value of longitudinal field. 
This provides a clear indication of replica symmetric or ergodic spin glass phase in presence of a small but finite value of longitudinal field supporting the observation of Schindler et al.~\ct{schindler22} as explained before.
We predict that this ergodicity is manifested in the dynamics of QA which is eventually responsible for the advantage of QA for $h$-annealed cases over the constant $h$ ones.

\section{Conclusions}
\label{conclu}
In summary, motivated by recent studies on the static properties of the quantum SK spin glass in presence of the longitudinal field~\ct{schindler22,kiss23}, we have revisited the problem of QA by tuning both transverse and longitudinal fields. Previously, we proposed~\cite{rajak14} this method of annealing (which, to
our knowledge, is the first report on such beneficial effect of
longitudinal field in quantum annealing) for a small system size of $N=8$. In this work, we have provided a detailed study of QA by tuning both transverse and longitudinal fields for three systems sizes $N=8, 10$ and $12$ for various disorder realizations. For almost all the cases, we have shown some advantages while longitudinal field is annealed rather than keeping it constant throughout the process by analyzing the overlap of the final state with the exact ground state(s) of the classical Hamiltonian. We have also observed that the difference between the final average probabilities for $h$-annealed and $h=0.1$ cases increases with the increase of the system size although their absolute values indeed decrease.

In search for the possible reason of the
enhanced quantum annealing
in presence of tuned longitudinal field (see
Figs.~\ref{fig_N8},\ref{fig_N10},\ref{fig_N12}), we have also investigated the
effect of non-vanishing longitudinal field
on the spin glass order parameter distribution
in the SK glass, using exact diagonalization
results for small system sizes (see Fig.~\ref{fig_op_dist}).
The usual SK model order parameter distribution
shows then a tendency to converge towards a
delta function  behavior (peaking to a single
order parameter value) in thermodynamic limit.
This indicates the disappearance of non-ergodic
(or replica symmetry broken) spin glass phase
in favor of an ergodic glass phase, helping
faster annealing (in presence of nonvanishing
longitudinal field) to the ground states, due
to quantum tunneling \ct{das08,rajak22,ray89}. These also
support the observation \ct{schindler22} of disappearance
of the replica symmetry broken spin glass
phase in the Almeida-Thouless region of the
SK spin glass. We hope to report soon
on a detailed study on these.

In brief,  even though our system size is
very small, we can clearly anticipate that
the order parameter distribution will converge
to a single value even for an arbitrarily small
longitudinal field in thermodynamically large
systems. This perhaps opens a new route to
effective quantum annealing (with fine tuning
or annealing of  the longitudinal field, along
with the transverse field), as we proposed
earlier \ct{rajak14}.

\section{Acknowledgements}
\label{ack}
%We are grateful to the guest
%editors Soumen Roy and Deepak Dhar for their
%invitation to contribute in this Special
%Issue on ``Statistical Physics and Complex Systems"
%of the Indian Journal of Physics, which started
%its journey in 1926, with Chandrasekhar Venkat
%Raman as its founder and first editor. 
AR is
thankful to UGC, India for Start-up Research 
Grant No. F. 30-509/2020(BSR) and BKC is thankful to
the Indian National Science Academy for their
Senior Scientist Research Grant support.

\end{document}